# Optical determination of the Néel vector in a CuMnAs thin-film antiferromagnet

V. Saidl[1], P. Němec[1,*], P. Wadley[2], V. Hills[2], R.P. Campion[2], V. Novák[3], K.W. Edmonds[2], F. Maccherozzi[4], S. S. Dhesi[4], B.L. Gallagher[2], F. Trojánek[1], J. Kuneš[3], J. Železný[3], P. Malý[1], and T. Jungwirth[3,2]

[1]Faculty of Mathematics and Physics, Charles University in Prague, Ke Karlovu 3, 121 16 Prague 2, Czech Republic
[2]School of Physics and Astronomy, University of Nottingham, Nottingham NG7 2RD, United Kingdom
[3]Institute of Physics ASCR, v.v.i., Cukrovarnická 10, 162 53 Prague 6, Czech Republic
[4]Diamond Light Source, Chilton, Didcot, Oxfordshire, OX11 0DE, United Kingdom

**Recent breakthroughs in electrical detection and manipulation of antiferromagnets have opened a new avenue in the research of non-volatile spintronic devices.[1-10] Antiparallel spin sublattices in antiferromagnets, producing zero dipolar fields, lead to the insensitivity to magnetic field perturbations, multi-level stability, ultra-fast spin dynamics and other favorable characteristics which may find utility in fields ranging from magnetic memories to optical signal processing. However, the absence of a net magnetic moment and the ultra-short magnetization dynamics timescales make antiferromagnets notoriously difficult to study by common magnetometers or magnetic resonance techniques. In this paper we demonstrate the experimental determination of the Néel vector in a thin film of antiferromagnetic CuMnAs[9,10] which is the prominent material used in the first realization of antiferromagnetic memory chips.[10] We employ a femtosecond pump-probe magneto-optical experiment based on magnetic linear dichroism. This table-top optical method is considerably more accessible than the traditionally employed large scale facility techniques like neutron diffraction[11] and X-ray magnetic dichroism measurements.[12-14] This optical technique allows an unambiguous direct determination of the Néel vector orientation in thin antiferromagnetic films utilized in devices directly from measured data without fitting to a theoretical model.**

Well-established optical methods[15] enable to study magnetic materials with a high spatial-resolution[16] on short time-scales.[17] In particular, Kerr and Faraday magneto-optical (MO) effects, which are linear in magnetization, are frequently used for the characterization of ferromagnets (FMs).[15-17] For antiferromagnets (AFs), the utilization of MO techniques is much more challenging. Several time-resolved studies have been performed on canted antiferromagnets[18-20] where the Dzyaloshinskii-Moriya interaction induces a canting of the two AF spin-sublattices with an angle of about 1° which leads to a small net magnetization. These canted AFs are, therefore, much easier for the experimental study because despite their antiferromagnetic ordering it is still possible to influence the spins with relatively weak magnetic fields and, moreover, Kerr and Faraday effects can be used for the characterization of their magnetic ordering. For fully compensated AFs the signals from oppositely oriented magnetic sublattices cancel for MO effects linear (i.e. odd) in magnetization which leaves only MO effects quadratic (even) in magnetization as suitable probes for these materials.[14] Quadratic MO effects have been reported for many magnetic materials.[15-17,21] However, practical utilization of quadratic MO effects for the characterization of magnetic materials is much less common than utilization of linear MO effects. [14,22,23] One reason for this is that quadratic MO effects are typically much weaker than linear MO effects.[14] Moreover, the experimentally measured light polarization change does not alter sign when the direction of magnetization in the sample is reversed (for example by external magnetic field in the case of FMs). Experimentally, especially the second feature is rather limiting because it significantly complicates the separation of the magnetic order-related signal from other sources of the light polarization change (e.g., strain- or crystal structure-related).[24] In this paper we show that this problem can be circumvented if a two-beam pump-probe detection scheme is used to measure the pump-induced demagnetization-related MO signal in the AF.

---

* Electronic mail: nemec@karlov.mff.cuni.cz

To demonstrate this technique we use a model collinear antiferromagnet, tetragonal CuMnAs[9] which is grown epitaxially on GaP(001). The spin axis lies in the *ab* plane of CuMnAs, which registers with the GaP substrate through a 45 degree rotation, making the CuMnAs{100} parallel to the GaP {110} (see Fig. 2 in Ref. 9).

In Fig. 1(a) we schematically illustrate the magnetic linear dichroism (MLD) in a compensated antiferromagnet. The rotation of light polarization (or the change of its ellipticity) occurs due to the different absorption coefficients for normal incidence light with a polarization plane oriented parallel and perpendicular to magnetic moments (see Fig. S1(a) and Eqs. (S1) and (S2) in the Supplementary material). In common magnetic materials, polarization rotation due to MLD does not exceed typically milliradians.[15-17,21] Consequently, it is rather difficult to separate experimentally this MO signal from other (non-magnetic) sources of light polarization change. In Fig. 1(b) we illustrate a situation when a second – considerably stronger – pump laser pulse illuminates the place where the probe pulse is experiencing the sample MO response. The pump-induced local heating of the sample leads to its partial demagnetization[25,26] and, consequently, to a reduction of the static MO signal. The MO signal measured in this pump-probe experiment is given by (see Eqs. (S3) and (S4) in the Supplementary material)

$$MO(\Delta t, \varepsilon) = \frac{2P^{MLD}}{M_S} sin2(\varphi - \varepsilon)\delta M_S(\Delta t) \quad , \tag{1}$$

where $P^{MLD}$ is the corresponding MO coefficient, which scales quadratically with the sublattice magnetization ($M_S$) projection onto the plane perpendicular to the probe light propagation direction. $\varphi$ and $\varepsilon$ describe the in-plane orientation of magnetic moments and light polarization, respectively (see Fig. S1(b) in the Supplementary material). In Figs. 1(c) and 1(d) we show the time-resolved pump-induced change of the MO signal and of the sample transmission, respectively, measured at 15 K in a transmission geometry in our 10 nm thin CuMnAs epilayer by probe pulses with several orientations of the polarization plane $\varepsilon$. In Fig. 1(e) we show a more detailed polarization dependence of the measured MO signal which confirms the harmonic dependence predicted by Eq. (1) for the MLD-related MO signal.

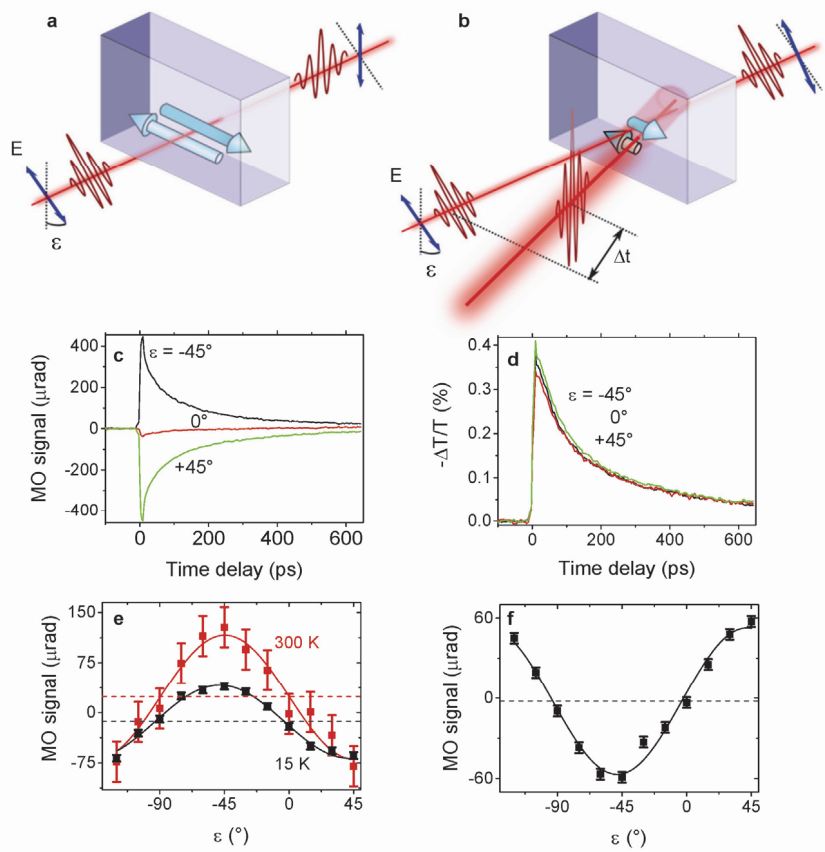

**Figure 1: Experimental observation of uniaxial magnetic anisotropy in 10 nm film of CuMnAs. a,** Schematic illustration of magnetic linear dichroism (MLD): different absorption of light polarized along and perpendicular to magnetic moments leads to a rotation of linearly polarized light polarization plane. **b,** Strong laser pulse-induced local heating of the sample leads to a demagnetization and, consequently, to a reduction of MLD-related probe polarization rotation, which is detected as a function of time delay $\Delta t$ between pump and probe pulses for various probe pulses polarization plane orientations $\varepsilon$. **c,** Time-resolved pump-induced change of MO signal measured by probe pulses with different $\varepsilon$ in transmission geometry; sample temperature 15 K, wavelength of pump and probe pulses 920 nm. **d,** Time-resolved transient transmission measured simultaneously with the MO signal shown in **c**. **e,** Probe-polarization dependence of pump-induced change of MO signal measured in transmission geometry for $\Delta t = 60$ ps at 15 K (black points) and 300 K (red points) using pump and probe pulses at 756 nm; $\varepsilon = 0°$ corresponds to polarization plane along the crystallographic direction [1-10] in GaP substrate, which coincides with the [100] direction in CuMnAs. Solid lines are fits by Eq. (1) plus polarization-independent backgrounds, which are shown by dashed horizontal lines. **f,** Same as **e** only in reflection geometry at 15 K using wavelength of pump and probe pulses of 920 nm.

We stress that the key ingredient which enables separation of light polarization rotation due to MLD from that of a non-magnetic origin (e.g., due to a strain in cryostat windows or a crystal structure of the GaP substrate) is the local modification of the magnetic order of the investigated CuMnAs epilayer by pump pulses which is consequently measured by probe pulses. Any polarization changes experienced by probe pulses during their propagation in the optical setup that are not modified by the pump pulses are not detected by this technique. Hence, it is sensitive only to changes occurring in the ≈ 100 μm region where the pump and probe beams spatially overlap. When the sample temperature is increased to 300 K the overall character of the measured polarization dependence does not change significantly – only the signal magnitude is larger at the higher temperature [see Fig. 1(e)]. In Fig. 1(f) we show the polarization dependence measured in the same sample at 15 K in reflection geometry. The signal shows the same periodicity as the signal measured in transmission geometry, but the sign of the measured MO signal is opposite for these two geometries due to the opposite sign of $P^{MLD}$.

Without knowing the actual sign of $P^{MLD}$ in the studied material it is not possible to deduce the absolute Néel vector orientation from these measurements. Nevertheless, the measured data imply [see Eq. (1)] that the spin axis in the studied 10 nm film of CuMnAs is either along the direction corresponding to $\varepsilon = 0°$ or $\varepsilon = -90°$. For the actual sample orientation in Fig. 1, these directions correspond to the crystallographic directions [1-10] and [110] in GaP substrate, respectively. Consequently, we can conclude that for the 10 nm film of CuMnAs there is an in-plane uniaxial magnetic anisotropy such that the Néel vector is preferentially oriented along the [100] or [010] crystallographic direction in CuMnAs.

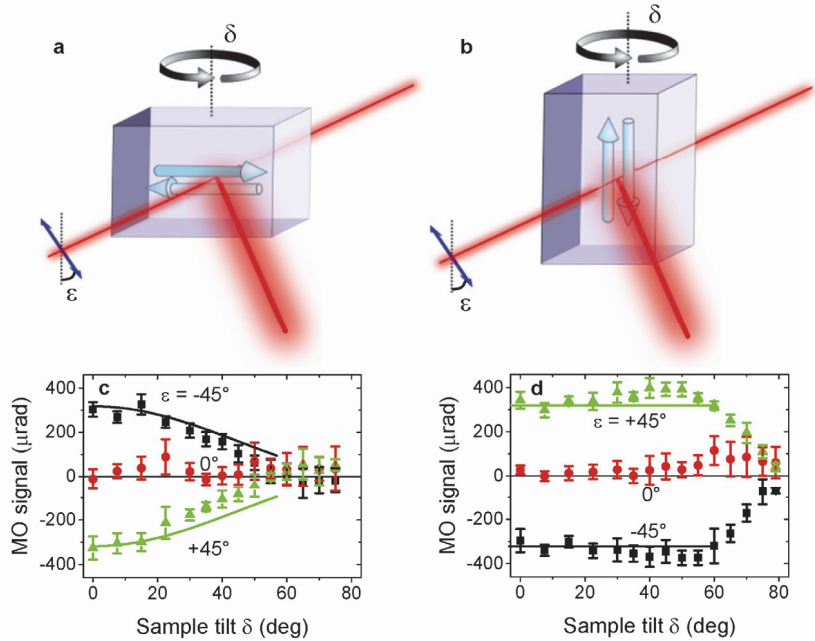

**Figure 2: Determination of the spin axis direction by MO experiment. a**, Schematic illustration of sample tilting around an axis that is perpendicular to the magnetic moments direction, which leads to a reduction of moments projection onto plane perpendicular to the probe light propagation direction. **b**, Schematic illustration of sample tilting around an axis that is parallel to the magnetic moments direction, which does not change the projection. **c**, Measured MO signal at 300 K for $\Delta t = 60$ ps (points) as a function of the sample tilt around the crystallographic direction [1-10] in GaP substrate (i.e., [100] direction in CuMnAs); wavelength of pump and probe pulses 920 nm. Solid lines depict function $cos^2\delta$ that describes the moment projection reduction expected for a sample tilt around the direction perpendicular to the spin axis (see also Eq. (S4) in the Supplementary material). **d**, Same as **c** for a sample tilt around the crystallographic direction [110] in GaP substrate (i.e., [010] direction in CuMnAs), which was identified as the spin axis direction, when the moment projection was not reduced by the sample tilt.

In Fig. 2 we demonstrate how the ambiguity in the Néel vector orientation can be removed on a purely experimental basis. As schematically illustrated in Figs. 2(a), tilting of the sample around an axis that is perpendicular to the magnetic moments direction leads to a reduction of the moment projection onto the plane perpendicular to the probe light propagation direction, i.e. to a reduction of $P^{MLD}$. On the other hand, the sample tilting around an axis that is parallel to the magnetic moment direction does not change this projection (see

Fig 2(b)). In Figs. 2(c) and 2(d) we show the measured MO signal as a function of the sample tilt $\delta$ around the GaP crystallographic directions [1-10] and [110], respectively. As expected, in one case the signal is reduced by the sample tilt [see Fig. 2(c)] while in the other case it is not affected significantly by the tilt up to $\delta \approx 60°$ [see Fig. 2(d)]. Above this value, the measured MO signal decreases which can be attributed to an increase of the illuminated spot size and/or a reduced spatial overlap between pump and probe pulses for these large incidence angles. Overall, we can conclude that from the performed MO experiments it follows that the spin axis direction in the 10 nm CuMnAs epilayer is along the crystallographic direction [110] of the GaP substrate that corresponds to the [010] direction in CuMnAs.

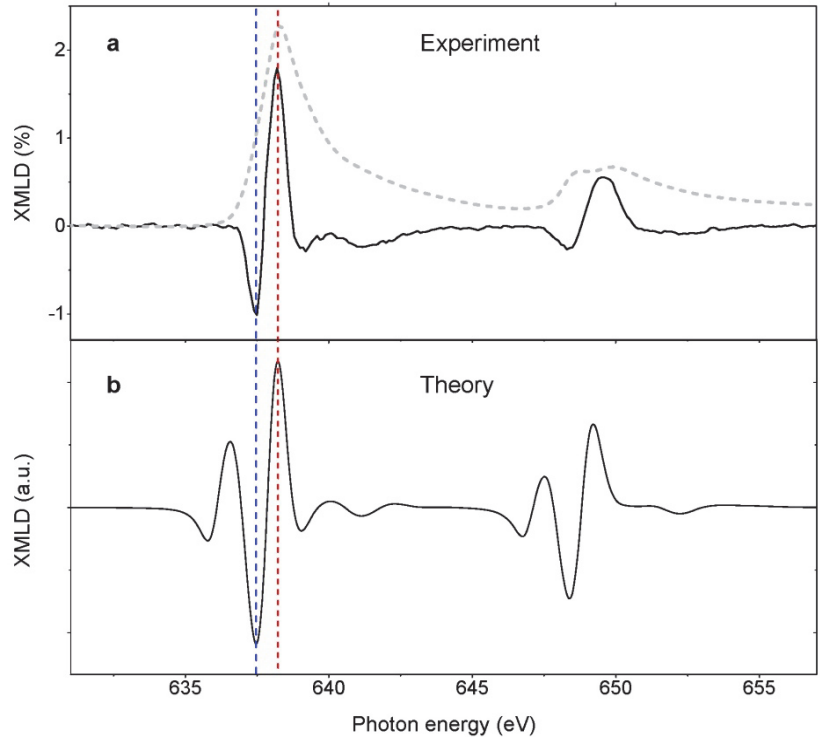

**Figure 3: Verification of the spin axis direction by XMLD. a**, Experimentally measured Mn $L_{2,3}$ XMLD spectrum, obtained for normal incidence x-rays as absorption for polarization along the [110] crystallographic direction in GaP minus the absorption for the polarization along the [1-10] direction. Red and blue dashed lines indicate the positions of the Mn $L_3$ absorption peak and valley. The absorption spectrum is represented by a dashed grey curve. **b**, Calculated XMLD spectrum for this geometry, defined as the absorption for AF moments parallel to the x-ray polarization, minus the absorption for AF moments perpendicular to the x-ray polarization.

To check the validity of this conclusion we performed a X-ray magnetic linear dichroism (XMLD) study. The XMLD was detected using total electron yield, with a probing depth of $\approx 3$ nm The surface of the uncapped 10 nm CuMnAs epilayer, which was used for the MO experiments described so far, is oxidized and, consequently, the measured XMLD signal was relatively weak in this sample. The surface oxidation was suppressed by using a 2 nm thick aluminium capping layer (we note we also see a clear uniaxial anisotropy in the MO signal in the Al-capped sample – see Supplementary Fig. S9). In Fig. 3(a) we show the XMLD spectrum measured in the capped 10 nm CuMnAs epilayer. The XMLD spectrum is obtained by taking the difference between Mn $L_{2,3}$ x-ray absorption spectra for x-rays at normal incidence with polarization parallel and perpendicular to the [110] direction of the GaP substrate. A clear linear dichroism is observed at the Mn $L_{2,3}$ edge while it is absent at the Cu $L_{2,3}$ edges,[27] which confirms the magnetic origin of the measured XMLD signal. In order to determine the spin axis orientation from this experiment, we compared the measured XMLD spectrum with that predicted by ab initio calculations.[28] In Fig. 3(b) we plot the calculated XMLD spectrum, defined as the absorption for AF moments parallel to the x-ray polarization, minus the absorption for AF moments perpendicular to the x-ray polarization. The positions of the principal peaks at both the $L_3$ and $L_2$ edges are in good agreement with the experimental data. From the XMLD sign we infer that in the 10 nm CuMnAs film the spin axis is along the crystallographic direction [110] of the GaP substrate, i.e., the [010] direction in CuMnAs, in agreement with the conclusions derived from the MO experiments described above. In a self-standing tetragonal crystal of CuMnAs, the directions [100] and [010] should be equivalent. However, it is the surface reconstruction of the zinc-blende GaP substrate, which determines the specific symmetry of bond alignments on the CuMnAs/GaP interface, that breaks the symmetry between these two crystal directions. Note that the present situation in CuMnAs/GaP is fully analogous to the deposition of Fe on a GaAs substrate where this surface-induced uniaxial magnetic anisotropy was investigated in detail.[29,30] Moreover, this interpretation of the origin of the in-plane uniaxial anisotropy in the 10 nm CuMnAs epilayer on the GaP substrate is also supported by the fact that we do not observe any in-plane uniaxial anisotropy in thicker films (see Supplementary Fig. S11) where, instead, the in-plane biaxial anisotropy of bulk CuMnAs dominates.[10]

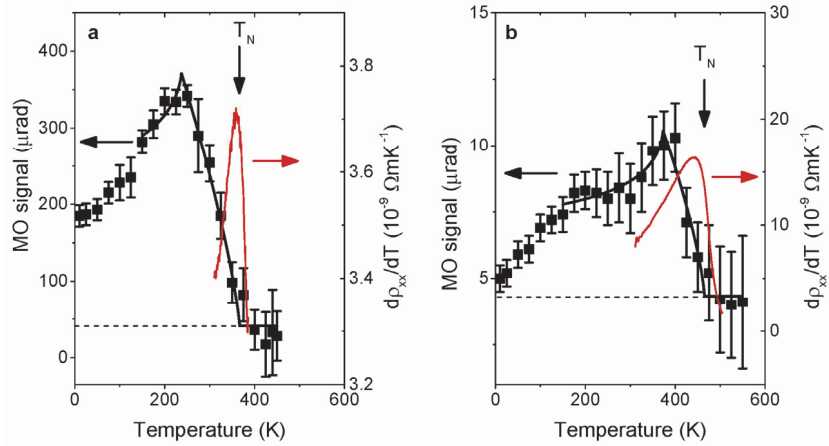

**Figure 4: Determination of the Néel temperature. a**, Temperature dependence of MO signal measured in uncapped 10 nm CuMnAs film in transmission geometry using pump and probe pulses at 920 nm (points). The black solid line is a fit by Supplementary Eq. (S10) plus background, which is shown by a dashed horizontal line, yielding a Néel temperature $T_N = (365 \pm 20)$ K, which is depicted by a vertical arrow. The red line is a temperature derivative of the sample resistivity. **b**, Same as **a** for 10 nm CuMnAs film with a protective Al layer. MO signal was measured in reflection geometry using pump pulses at 810 nm and probe pulses at 405 nm; the deduced Néel temperature $T_N = (465 \pm 20)$ K.

In Fig. 4 we show the measured temperature dependence of the MO signal from which the Néel temperature can be determined. Without the pump pulse, the temperature dependence of the quadratic MO signal can be approximated[27] by a power law $(T_N - T)^{2\beta}$, where $T$ is the sample temperature, $T_N$ is the Néel temperature and $\beta$ is the critical exponent (see the Supplementary material for more details). Absorption of the pump pulse leads to a sample temperature increase $\Delta T$ which results in a change of the sample MO activity, which is the signal measured in the pump-probe experiment (see the Supplementary Fig. S8). By fitting the measured MO data for the uncapped CuMnAs epilayer [see Fig. 4(a)] we obtained a Néel temperature $T_N = (365 \pm 20)$ K and a pump-induced temperature increase $\Delta T = (120 \pm 30)$ K. The deduced Néel temperature agrees well with that obtained[31] from the peak position in the temperature derivative of the sample resistivity, which is shown as the red solid line in Fig. 4(a). The CuMnAs epilayer capped with aluminium has a considerable higher Néel temperature of $T_N = (465 \pm 20)$ K [see Fig. 4(b)], which is close to the value $T_N = (480 \pm 5)$ K reported previously for a 500 nm thick CuMnAs film.[27]

In summary, we have demonstrated the optical determination of the Néel vector orientation and Néel temperature in a fully compensated AF thin film of CuMnAs. We have utilized magnetic linear dichroism combined with pump-probe experimental technique of ultrafast laser spectroscopy. The research potential of the technique was demonstrated with measurements on CuMnAs which is a promising material for the development of antiferromagnetic memories. We emphasize that quadratic magneto-optical effects employed in our experiments are generic in antiferromagnets and, therefore, our approach should be applicable to a broad range of other material candidates for AF spintronics. As a table-top experiment, our approach is considerably more accessible than the large-facility measurements and it also provides a uniquely direct experimental means for detecting the Néel vector orientation in thin-film AFs. Intrinsically, our optical measurements provide the high time-resolution matching to the THz internal dynamics-scale of AFs, which is beyond the reach of the electrical detection tools demonstrated earlier.

**Methods**

**MO experiment.** A Ti:sapphire oscillator (Mai Tai, Spectra Physics) was used as a light source. Laser pulses, with a time width of 150 fs and repetition rate of 80 MHz, were divided into pump pulses (with a fluence of $\approx 3$ mJ/cm$^2$) and probe pulses (with at least 50 times weaker fluence) that were focused to a same spot on the sample (with a full width at half maximum, FWHM, of about 30 μm). Unless explicitly mentioned (i.e., except for Fig. 2), the experiments were performed close to the normal incidence geometry (with an angle between pump and probe beams of 6°). The polarization of the pump pulses was either circular or linear (see Supplementary Fig. S5 where the independence of the measured MO signal on the pump polarization is demonstrated), while the probe pulses were linearly polarized with the polarization plane orientation controlled by a half-wave plate and described by an angle $\varepsilon$ (see Supplementary Fig. S1(b) for a definition of the coordinate system). The dynamics of the polarization plane rotation for the transmitted or reflected probe pulses was obtained by taking a difference of signals measured by detectors in an optical bridge detection system.[32] Alternatively, the probe ellipticity was measured. The same dynamics of the pump-induced change of rotation and ellipticity for $\Delta t \geq 20$ ps (see Supplementary Fig. S4) confirms the magnetic origin of the measured MO signals[33] for $\Delta t \approx 60$ ps, which were used in the above analysis. Simultaneously with the MO signal we measured also the sum of signals from the detectors,[32] which corresponded to a probe intensity change due to the pump-induced modification of the sample transmission (or reflectivity). As a measure of the transmission changes we used the differential transmission $\Delta T/T = (T_E-T)/T$, where $T_E$ ($T$) is the transmission with (without) the pump pulse.[34] The very similar dynamics measured for MO signal and $\Delta T/T$ [see Figs. 1(a) and 1(b)] is a strong indication that both these signals are due to the pump-induced temperature increase.[35] We have also verified that the measured probe-polarization dependence of the MO signal is connected with the orientation of the sample (see Supplementary Fig. S2) and that the MO signal measured in the bare GaP substrate is considerably weaker (see Supplementary Fig. S3) than those measured in samples with CuMnAs epilayers. The sample was mounted on a cold finger of the closed-cycle helium cryostat (ARS) where the temperature can be changed from 15 to 800 K. The experiments were performed without external magnetic field applied because control experiments showed that the external magnetic field up to 500 mT (generated by an electromagnet in the sample plane) has no effect on the measured data (see Supplementary Fig. S6). The majority of pump-probe experiments was performed for a same wavelength of pump and probe pulses – see Supplementary Fig. S7 for the measured spectral dependence of the MO signal. However, for the Al-capped sample also experiments using pump pulses at 810 nm and probe pulses at 405 nm (second harmonic generated in a BBO crystal) were performed.

**XMLD experiment.** The XMLD study was performed on beamline I06 at Diamond Light Source. The variable polarization undulator of the beamline allows the x-ray linear polarization vector to be rotated by 90°, allowing XMLD measurements without moving the sample. Mn and Cu $L_{2,3}$ x-ray absorption edge spectra were obtained using the total electron yield method, with x-rays at normal incidence. The experiment was performed at a temperature of $T = 250$ K.

**Acknowledgements**

This work was supported by the Grant Agency of the Czech Republic under Grant No. 14-37427G, by the EU ERC Advanced Grant No. 268066, by the Ministry of Education of the Czech Republic Grant No. LM2011026, by the University of Nottingham EPSRC Impact Acceleration Account, and by the Grant Agency of Charles University in Prague Grants no. 1910214 and SVV–2015–260216. We thank Diamond Light Source for the provision of beamtime under proposal number SI-9993.

**Author contributions**

R.C., V.H. and V.N. prepared the samples, P.N., P.W, B.G., P.M. and T.J. planned the experiments, V.S. and F.T. performed the MO experiments, V.H. performed the electrical measurements, P.W., K.W.E., F.M. and S.D. performed the XMLD experiment, J.K. , J.Z. performed the XMLD calculations, P.N., V.S. and T.J. wrote the manuscript with contributions from all authors.

**Additional information**

Supplementary information is available in the online version of the paper. Correspondence and requests for materials should be addressed to P.N.

# Optical determination of the Néel vector in a CuMnAs thin-film antiferromagnet: Supplementary information

V. Saidl[1], P. Němec[1], P. Wadley[2], V. Hills[2], R.P. Campion[2], V. Novák[3], K.W. Edmonds[2], F. Maccherozzi[4], S. S. Dhesi[4], B.L. Gallagher[2], F. Trojánek[1], J. Kuneš[3], J. Železný[3], P. Malý[1], and T. Jungwirth[3,2]

[1]Faculty of Mathematics and Physics, Charles University in Prague, Ke Karlovu 3, 121 16 Prague 2, Czech Republic
[2]School of Physics and Astronomy, University of Nottingham, Nottingham NG7 2RD, United Kingdom
[3]Institute of Physics ASCR, v.v.i., Cukrovarnická 10, 162 53 Prague 6, Czech Republic
[4]Diamond Light Source, Chilton, Didcot, Oxfordshire OX11 0DE, United Kingdom.

This supplementary material provides detailed additional information about time-resolved magneto-optical (MO) experiments in CuMnAs epitaxial layers.

## Magnetic linear dichroism

We investigated properties of CuMnAs epilayers by a pump-and-probe experimental technique (details are provided in the Methods section of the main text). The rotation of the polarization plane of probe pulses is caused by magnetic linear dichroism (MLD).[s1] This MO effect is schematically illustrated in Fig. S1a. Let us suppose that linearly polarized light arrives under normal incidence to the surface of the (anti)ferromagnetic sample with in-plane orientation of magnetic moments. The light polarization, described by an electrical vector $E$, can be written as a sum of two mutually orthogonal polarizations: $E_{//}$ which is parallel to the magnetic moments and $E_{\perp}$ which is perpendicular to the magnetic moments. Behind the sample the electrical vectors are $E'_{//} = aE_{//}$ and $E'_{\perp} = bE_{\perp}$ where $a$ and $b$ are the corresponding amplitude transmission coefficients. Due to MLD light with one polarization is absorbed more in the sample than light with the second polarization, i.e., $a \neq b$. Consequently, the orientation of $E'$ behind the sample is rotated by an angle $\theta$ with respect to $E$. If we assume that $a/b \approx 1$ (i.e., that $\theta$ is small) we obtain [s1]

$$MO^{stat}(\varepsilon) \equiv \theta = P^{MLD} \sin 2(\varphi - \varepsilon), \tag{S1}$$

where $\varphi$ and $\varepsilon$ describe the in-plane orientation of magnetic moments and light polarization, respectively [see Fig. S1(b)], and the magneto-optical coefficient $P^{MLD}$ is defined as

$$P^{MLD} = 0.5\left(\frac{a}{b} - 1\right). \tag{S2}$$

However, it is usually quite difficult to measure this static MO signal due to contributions from other effects. The solution of this problem is to heat up the sample locally by a pump pulse which decreases the (sublattice) magnetization $M_S$. Consequently, in the excited region the polarization rotation experienced by probe pulse, which depends through $P^{MLD}$ quadratically on $M_S$, is smaller in a comparison with the situation without the pump pulse (see Fig. 1(b) in the main text). This polarization rotation change caused by a (relatively small) pump-induced demagnetization is

$$\delta MO(\Delta t, \varepsilon) \equiv MO^{stat}(\varepsilon) - MO^{with\ pump}(\Delta t, \varepsilon) = \frac{2P^{MLD}}{M_S} \sin 2(\varphi - \varepsilon)\delta M_S(\Delta t), \quad (S3)$$

where $\delta M_S(\Delta t)$ is the pump-induced change of $M_S$.

In a more general case, when light arrives with an angle of incidence $\delta$ (see Fig. 2(a) in the main text), the rotation of light polarization by MLD is given by

$$\delta MO(\Delta t, \varepsilon) = \frac{2P^{MLD}}{M_S} \cos^2 \delta \sin 2(\varphi - \varepsilon)\delta M_S(\Delta t) \ . \qquad (S4)$$

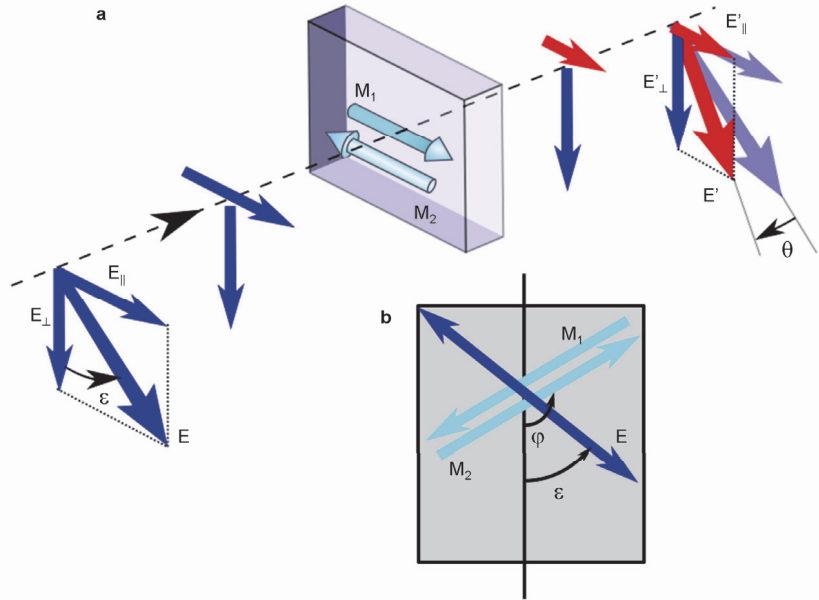

**Figure S1: Schematic illustration of magnetic linear dichroism.** (a) Linearly polarized light can be written as a sum of two perpendicular polarizations which are parallel and perpendicular to the magnetic moments (depicted by sublattice magnetizations $M_1$ and $M_2$), respectively. These two polarizations are absorbed differently and, consequently, the light polarization plane is rotated by angle $\theta$ behind the sample. (b) For light at normal incidence, magnetic linear dichroism is sensitive to a relative orientation of the in-plane projections of $M_1$ and $M_2$, which are described by angle $\varphi$, and light polarization plane, which is described by angle $\theta$. The angles are positive for counterclockwise direction when the sample is observed along the incident light; $\varepsilon = 0°$ correspond to *s*-polarized (vertical) linear polarization.

## Verification of CuMnAs origin of measured MO data

In Fig. 1 in the main text we show that the detected MO signals depend on the polarization of the probe pulses, as predicted by Eq. (S3). To confirm that this signal is really connected with the easy axis position in the CuMnAs layer, we performed measurements for two orientations of the sample, which differ by 45° in-plane rotation (see inset in Fig. S2). In Fig. S2 we illustrate that for a fixed probe polarization orientation ($\varepsilon$) the rotation of the sample (i.e., the easy axis position, $\varphi$) by 45° leads exactly to the behavior predicted by Eq. (S3).

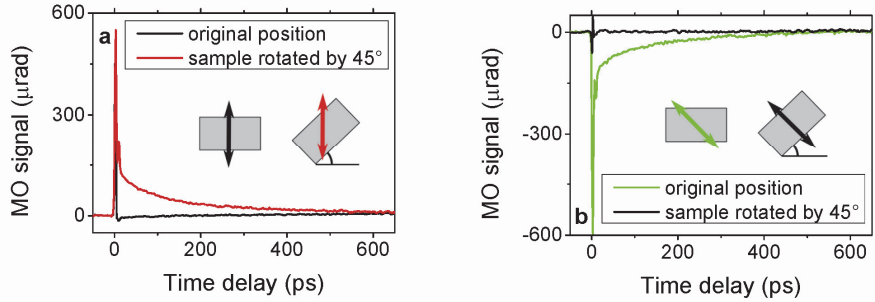

**Figure S2: Verification of MO signal change for rotated sample.** Comparison of MO signals for two orientations of the sample, which differ from each other by in-plane rotation for 45°, measured using (a) vertical ($\varepsilon = 0°$) and (b) diagonal ($\varepsilon = 45°$) probe polarizations, as depicted in insets. Data were measured in transmission geometry in uncapped 10 nm CuMnAs film at 15 K using pump and probe pulses at 820 nm.

Theoretically, it might be also possible that the measured MO signals come from the GaP substrate and not from the CuMnAs layer. To verify this, we performed the experiment also in a bare GaP substrate without any CuMnAs layer – the obtained results are shown in Fig. S3. Clearly, the detected MO signal is much weaker and, moreover, there is not any pronounced polarization dependence. Therefore, the experiments shown in Figs. S2 and S3 clearly confirm that the MO signals shown in Fig. 1 in the main paper come from the CuMnAs epilayer.

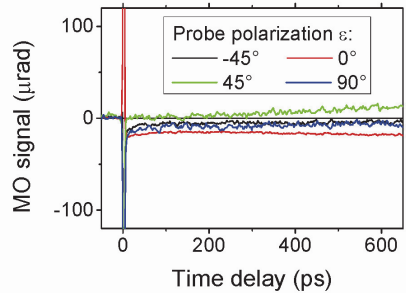

**Figure S3: MO signal measured in bare GaP substrate** for several orientations of probe pulses polarization plane. Data were measured in transmission geometry at 15 K using pump and probe pulses at 820 nm.

**Verification of magnetic origin of measured MO data**

In principle, the measured dynamical MO signals can contain information not only about the pump-induced magnetization change but also about the pump-induced change of the complex index of refraction (the so-called "optical part" of the MO signal).[s2] Consequently, the dynamics of both the rotation and ellipticity has to be measured and compared before the obtained magneto-optical signal is attributed to the magnetization dynamics.[s2] The polarization rotation ($\theta$) and elipticity ($\eta$) induced by MLD can be described as

$$\theta \approx f_\theta M^2, \tag{S5}$$
$$\eta \approx f_\eta M^2. \tag{S6}$$

Here $M$ is the magnetization and $f_\theta$ and $f_\eta$ are functions that depend on the electronic properties of the material and that can be expressed in terms of the refractive index and the absorption coefficient. Consequently, the pump-induced change of rotation $\delta\theta(t)$ and elipticity $\delta\eta(t)$ can be written as

$$\delta\theta(\Delta t) \approx f_\theta 2M\delta M(\Delta t) + \delta f_\theta(\Delta t)M^2, \tag{S7}$$
$$\delta\eta(\Delta t) \approx f_\eta 2M\delta M(\Delta t) + \delta f_\eta(\Delta t)M^2. \tag{S8}$$

These formulas show that if the change of the magnetization $\delta M(\Delta t)$ is the main contribution in the measured MO data, the pump-induced dynamics of rotation and elipticity should have the same time dependence. They should differ only in a magnitude because of a different constants $f_\theta$ and $f_\eta$. On the contrary, if the pump-induced change of optical properties is the main contribution to the measured signal, the dynamics of rotation and elipticity cannot be the same because $\delta f_\theta(\Delta t)$ and $\delta f_\eta(\Delta t)$ are generally different. In Fig. S4 we show the experimentally measured pump-induced change of the rotation and elipticity. The magnitude of the elipticity signal is approximately 2.4-times larger than that of the rotation, but their shapes are the same for $\Delta t \geq 20$ ps. This confirms that the measured MO data have a magnetic origin.

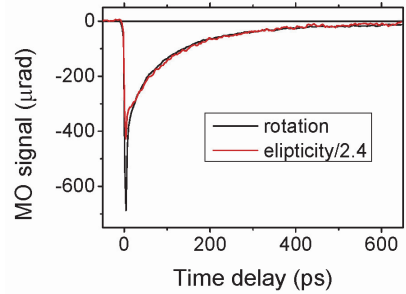

**Figure S4: Comparison of pump-induced rotation and elipticity change**. Data were measured in transmission geometry in uncapped 10 nm CuMnAs film at 15 K using pump and probe pulses at 920 nm.

### Independence of measured MO signals on pump pulse polarization

In Fig. S5 we illustrate that the measured MO signals do not depend on the polarization of pump pulses. This is in accord with our interpretation that the measured MO signal originates from the pump-induced thermal demagnetization.

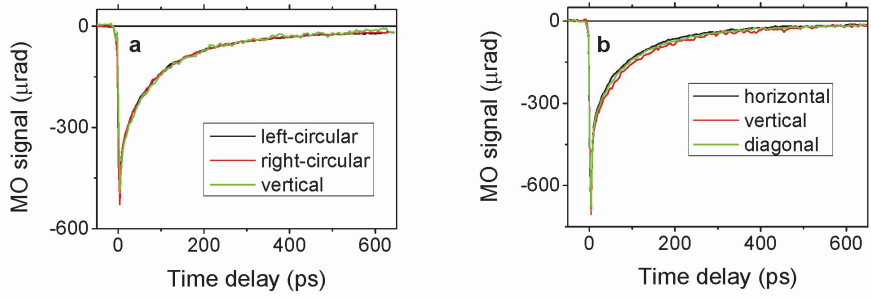

**Figure S5: Independence of MO signal on pump pulse polarization. a**, MO signal measured for circularly (left and right) and linearly (vertically) polarized pump pulses. **b**, MO signal measured for linearly polarized pump pulses. Data were measured in transmission geometry in uncapped 10 nm CuMnAs film at 15 K using pump and probe pulses at 820 nm.

### Independence of measured MO signal on external magnetic field

Magnetic properties of the CuMnAs layers were examined using a neutron diffraction and superconducting quantum interference device (SQUID) and it was shown that this material is a fully compensated antiferromagnet.[s3] To verify that the measured MO signals are not, for example, due to some uncompensated magnetic moments present in the studied samples, we performed the experiments with and without an external magnetic field of 500 mT. In Fig. S6 we show that this magnetic field has absolutely no effect on the measured MO signal, which is the expected result for a compensated antiferromagnet.

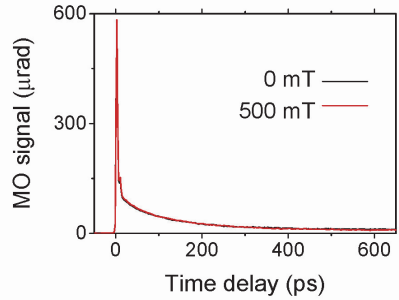

**Figure S6: Independence of MO signal on external magnetic field.** Magnetic field was applied at 45° from the sample easy axis. Data were measured in transmission geometry in uncapped 10 nm CuMnAs film at 15 K using pump and probe pulses at 820 nm.

### Spectral dependence of measured MO signal

It is quite difficult to measure a sign and a magnitude of static MO coefficient $P^{MLD}$ in compensated antiferromagnets. However, we can deduce a spectral dependence of the relative magnitude of $P^{MLD}$ from a spectral dependence of the measured dynamical MO signal, which is shown in Fig. S7. Pump and probe pulses had the same wavelength in this degenerate experiment. Absorption coefficient of CuMnAs almost does not depend on the wavelength from 700 nm to 1000 nm, where the Ti:Al$_2$O$_3$ laser is tunable. Consequently, for a constant fluence of pump pulses, the pump-induced demagnetization is the same for all the wavelengths from this spectra range and, therefore, the measured spectral dependence shown in Fig. S7 is given solely by a spectral dependence of $P^{MLD}$ [cf. Eq. (S3)]. From the measured dependence of the MO signal on the probe polarization (see Fig. 1 in the main text) and the known orientation of the spin axis (which is along the crystallographic direction [110] of the GaP substrate) and a negative sign of $\delta M_S$ we can also deduce that in the used transmission geometry $P^{MLD} > 0$ in the spectral range from 720 nm to 920 nm.

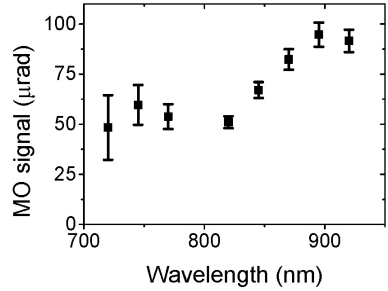

**Figure S7: Spectral dependence of MO signal** for time delay $\Delta t = 60$ ps. Data were measured in transmission geometry in uncapped 10 nm CuMnAs film at 15 K using the same wavelength of pump and probe pulses and a constant fluence of pump pulses.

### Determination of Néel temperature from measured MO signals

MLD is MO effect quadratic in magnetization. Therefore, unlike for Kerr and Faraday effects, the MO signals from oppositely oriented magnetic sublattices add up. For each sublattice, the temperature dependence of magnetization can be approximated near (and below) the phase transition temperature by a formula[s4]

$$M(T) \propto (T_N - T)^\beta, \qquad (S9)$$

where $T_N$ is a Néel temperature and $\beta$ is a critical exponent, which is for the three-dimensional Heisenberg model equal to 0.367.[s4-s6] Absorption of the pump pulse leads to a sample temperature increase $\Delta T$ which results in a modified temperature dependence of the sublattice magnetization, as schematically illustrated in Fig. S8(a). Therefore, the temperature dependence of the pump-induced change of the MO signal is given by

$$\text{for } T < T_N - \Delta T: \quad MO = A\left[(T_N - T)^{2\beta} - (T_N - T - \Delta T)^{2\beta}\right] \qquad (S10a)$$
$$T_N - \Delta T \leq T < T_N: \quad MO = A(T_N - T)^{2\beta} \qquad (S10b)$$
$$T \geq T_N: \quad MO = 0, \qquad (S10c)$$

where $A$ is the proportionality constant related to $P^{MLD}$. The measured MO data for the uncapped (Fig. 4(a) in the main text) and Al-capped (Fig. 4(b) in the main text) 10 nm CuMnAs epilayer were fitted by Eq. (S10) plus a temperature-independent background, which is presumably coming from the GaP substrate (see Fig. S3). The deduced critical exponent determined by the fitting $\beta = (0.37 \pm 0.03)$ agrees well with that predicted by the Heisenberg model.[s4-s6]

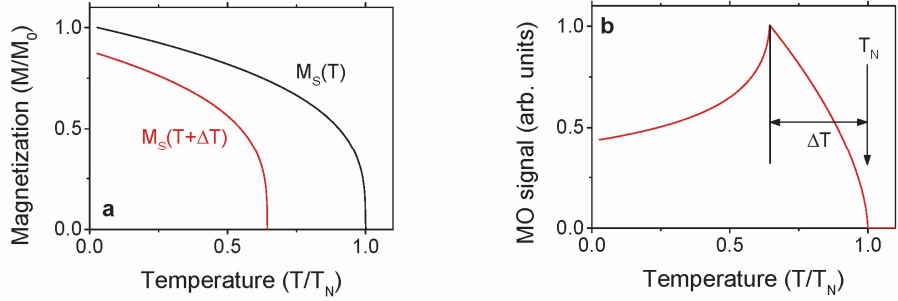

**Figure S8: Schematic illustration of effect of pump pulse-induced sample heating. a**, Temperature dependence of a sublattice magnetization $M_S$, which is described by Eq. (S9), without (black) and with (red) pump pulse, which leads to a sample temperature increase $\Delta T$. **b,** Expected temperature dependence of pump-induced MO signal, which is described by Eq. (S10).

We stress that Eq. (S9) describes correctly the magnetization magnitude only for temperatures close to the critical temperature. Therefore, these figures serve solely as schematic illustrations of the involved physics.

## Signal oscillations due to acoustic phonons

Time-resolved signals measured in reflection geometry (both the differential reflectivity and MO signal) differ from those shown for transmission geometry in Fig. 1(c) and (d) in the main text by a presence of additional oscillations – see Fig. S9(a) for data measured in Al-capped 10 nm CuMnAs film. Below we show that these oscillations are due to a propagation of coherent acoustic phonons in the GaP substrate[s7,s8] and that they are just added to the signals which are connected with the pump-induced demagnetization in CuMnAs. The role of this "parasitic" signal is negligible if the measured data are processed at sufficiently long time delays (e.g., for $\Delta t = 130$ ps in the case of Fig. S9), as illustrated in Fig. S9(b).

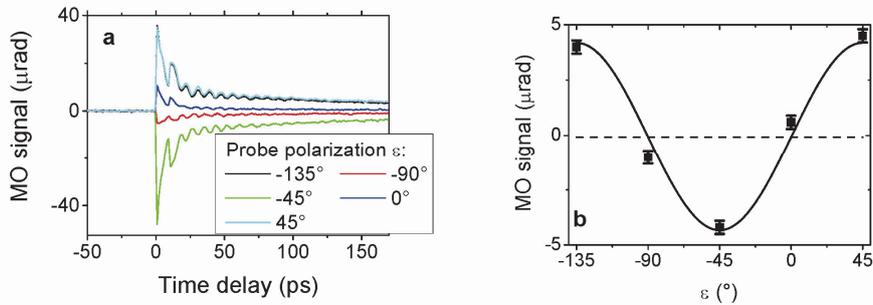

**Figure S9: Polarization dependence of MO signal in Al-capped 10 nm CuMnAs film. a**, Dependence of MO signal on time delay for several orientations of probe pulses polarization plane; the oscillation are due to propagation of coherent acoustic phonons. **b**, Dependence of MO signal measured for time delay 130 ps on probe polarization. Data were measured in reflection geometry at 15 K using pump pulses at 810 nm and probe pulses at 405 nm.

Absorption of the pump pulse in the CuMnAs layer generates a coherent wavepacket of accoustic phonons, which has the thickness of the excited CuMnAs layer and propagates into the GaP substrate.[s7] This leads to an appearance of a thin strained layer of GaP whose refractive index differs a little bit from that of the surrounding material and which moves deeper into the substrate with the acoustic sound velocity [see Fig. S10(a)]. The sample thus behaves as a Fabry-Perot interferometer and the oscillations in the optical signals are observed. The key fingerprint of this oscillation mechanism is the linear dependence[s7] of the oscillation period $\tau$ on the probe light wavelength $\lambda$

$$\tau = \frac{\lambda}{2v_s n(\lambda)}, \qquad (S11)$$

where $v_s$ is longitudinal acoustic sound velocity and $n(\lambda)$ is the wavelength-dependent refractive index, which is exactly the case for the data shown in Fig. S10(c).

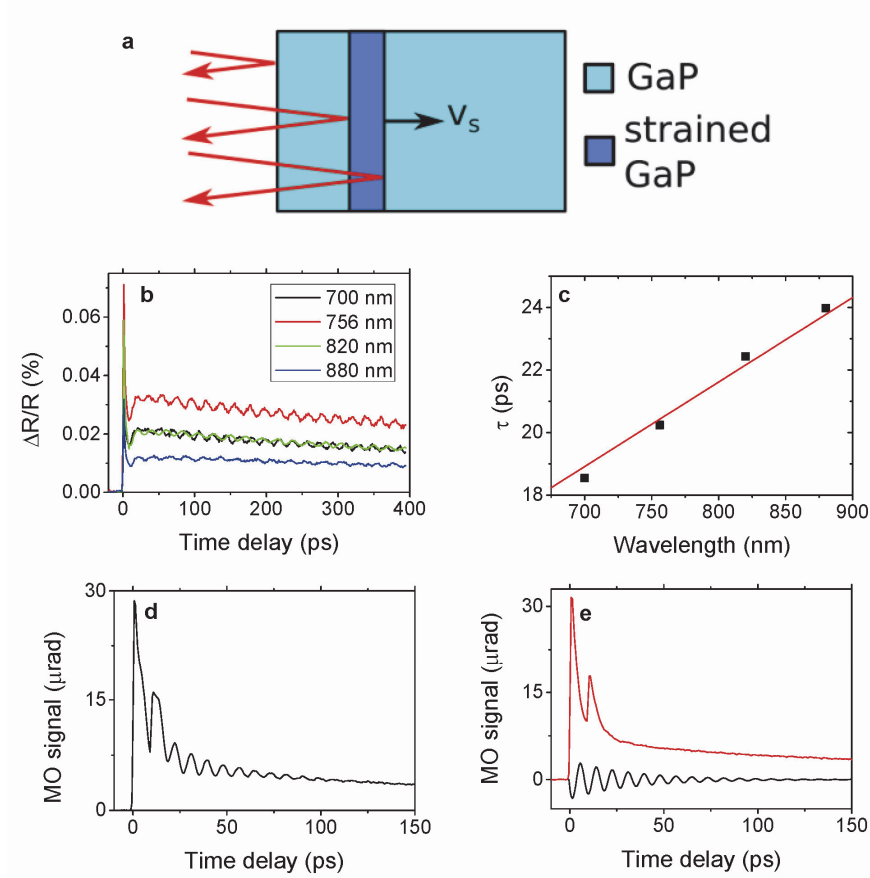

**Figure S10: Signal oscillations due to propagation of coherent acoustic phonons. a**, Schematic illustration of propagating strained GaP layer, which has a slightly different index of refraction and on which the light is partially reflected; $v_s$ is the longitudinal acoustic sound velocity. **b**, Dependence of differential reflectivity on time delay for various wavelengths. Data were measured in uncapped 46 nm CuMnAs film at 15 K using the same wavelength of pump and probe pulses for pump fluence 0.7 mJ/cm². **c**, Dependence of the oscillation period, which was deduced from part **b**, on laser wavelength (points). The line is a fit by Eq. (S11). **d**, As-measured time-resolved MO signal in Al-capped 10 nm CuMnAs film in reflection geometry at 15 K using pump pulses at 810 nm and probe pulses at 405 nm. **e**, Separation of the oscillatory signal due to acoustic phonons (black line) from the non-oscillatory MO signal (red line) for data shown in part **d**; the second peak in the non-oscillatory MO signal for time delay ≈10 ps is due to the pump pulse reflection from the substrate back side.

One possibility how the influence of this additional signal can be removed from the measured MO signal is to process the data at sufficiently long time delays where these oscillations are no longer visible, as illustrated in Fig. S9(b) above. Alternatively, we can fit the as-measured data by a damped harmonic function, which corresponds to this propagating coherent wavepacket of acoustic phonons, and numerically remove the oscillatory signal from the measured data. This procedure, which is illustrated in Figs. S10(d) and (e), was applied to obtain the data shown in Fig. 4(b) in the main text.

## MO signals measured in thicker CuMnAs epilayers

In thicker CuMnAs layers we do not see any significant pump-induced MO signals, as illustrated in Figs. S11(a) and (b) for 46 nm thick layer and in Figs. S11(c) and (d) for 128 nm thick layer. The most probable explanation of this fact is the existence of an in-plane biaxial anisotropy and a sub-micrometer size of domains, which are considerably smaller than the beam size, as was observed experimentally by XMLD-PEEM in the thicker films.[s9]

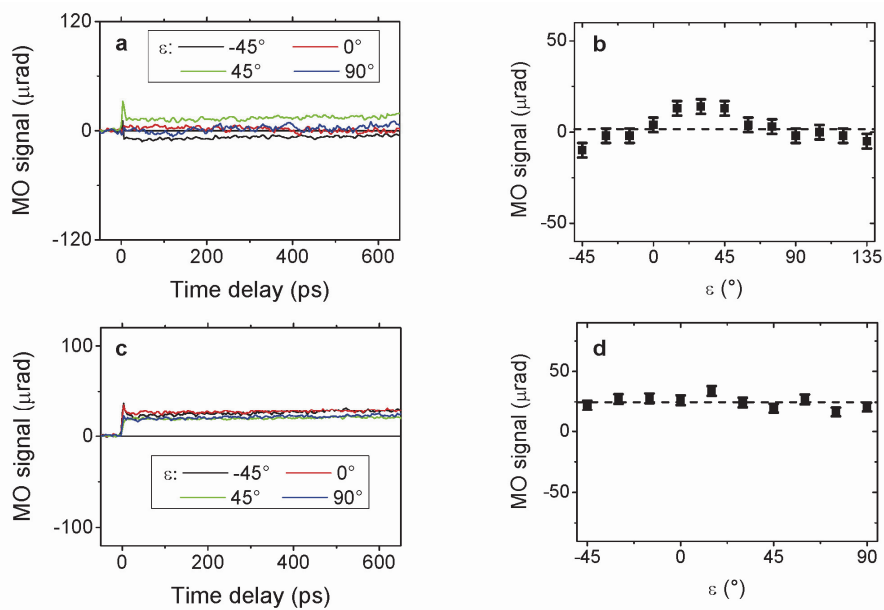

**Figure S11: Measured MO signals in thicker CuMnAs films**. **a**, Time-resolved MO signal in uncapped 46 nm thick layer for several orientations of probe pulses polarization plane. **b**, Probe polarization dependence of MO signal for uncapped 46 nm thick layer. **c,** and **d**, Same as **a,** and **b**, for uncapped 128 nm thick layer. Data were measured in transmission geometry at 15 K using pump and probe pulses at 820 nm.

**References**
[s1] Tesarova, N. *et al.* Direct measurement of the three-dimensional magnetization vector trajectory in GaMnAs by a magneto-optical pump-and-probe method, *Appl. Phys. Lett*. **100**, 102403 (2012).
[s2] Koopmans, B. *et al.* Ultrafast magneto-optics in nickel: magnetism or optics? *Phys. Rev. Lett.* **85**, 844-847 (2000).
[s3] Wadley, P. et al. Tetragonal phase of epitaxial room-temperature antiferromagnet CuMnAs. *Nature Commun.* 4, 2322 (2013).
[s4] Blundel, S. Magnetism in condensed matter (Oxford University Press, Oxford, 2004).
[s5] Jongh, L. J., Miedema, A. R. Experiments on simple magnetic model systems, *Advances in Physics* 50, 947-1170 (2001).
[s6] Guillou J. C., Zinn-Justin J. Critical exponents for the n-vector model in three dimensions from field theory, *Phys. Rev. Lett.* 39, 95 (1977).